# Optical hyperpolarization and NMR detection of $^{129}$Xe on a microfluidic chip

Ricardo Jiménez-Martínez[*,†], Daniel J. Kennedy[‡,§], Michael Rosenbluh[¶], Elizabeth A. Donley[*], Svenja Knappe[*], Scott J. Seltzer[‡,§], Hattie L. Ring[‡,§], Vikram S. Bajaj[‡,§,♦], and John Kitching[*,♦]

**Optically hyperpolarized $^{129}$Xe gas[1] has become a powerful contrast agent in nuclear magnetic resonance (NMR) spectroscopy and imaging, with applications[2] ranging from studies of the human lung[3] to the targeted detection of biomolecules[4,5]. Equally attractive is its potential use to enhance the sensitivity of microfluidic[6] NMR experiments, in which small sample volumes yield poor sensitivity. Unfortunately, most $^{129}$Xe polarization systems are large and non-portable. Here we present a microfabricated chip that optically polarizes $^{129}$Xe gas. We have achieved $^{129}$Xe polarizations greater than 0.5% at flow rates of several microliters per second, compatible with typical microfluidic applications. We employ *in situ* optical magnetometry[7] to sensitively detect and characterize the $^{129}$Xe polarization at magnetic fields of 1 µT. We construct the device using standard microfabrication techniques[8], which will facilitate its integration with existing microfluidic platforms. This device may enable the implementation of highly sensitive $^{129}$Xe NMR in compact, low-cost, portable devices.**

[*] National Institute of Standards and Technology, 325 Broadway, Boulder, CO 80305 USA, [†]Department of Physics, University of Colorado at Boulder, Boulder, CO 80309, USA, [‡]Materials Science Division, Lawrence Berkeley National Laboratory, Berkeley, California 94720-7300, USA, [§]Department of Chemistry, University of California at Berkeley, Berkeley, CA 94720, USA, [¶] The Jack and Pearl Resnick Institute for Advanced Technology, Department of Physics, Bar-Ilan University, Ramat-Gan IL-52900, Israel. [♦] e-mail:vsbajaj@lbl.gov; john.kitching@nist.gov

Most [129]Xe optical polarizers have been designed to produce litre-sized volumes of polarized gas[9,10]. These systems, designed for clinical magnetic resonance imaging or the large-scale imaging of materials, are necessarily bulky, expensive, and non-portable. Here, we present the fabrication and operation of a microfabricated source of hyperpolarized [129]Xe, suitable for the low-cost, power-efficient production of polarized gases and integration with microfluidic systems[6,11,12]. Our device uses an effective pumping volume of 25 µL and 8 mW of pumping light to achieve [129]Xe polarizations greater than 0.5% at gas flow rates of 5 µL/s; the mechanism of [129]Xe hyperpolarization occurs via spin-exchange optical pumping (SEOP) with [87]Rb atoms[1], in the absence of any superconducting magnet. This represents a signal enhancement of ~200 over the room temperature thermal equilibrium polarizations achievable at the highest commercially available magnetic field of 23.5 T. Further, our device requires only standard microfabrication techniques and benefits from some of the known strengths of this technology[13] such as batch fabrication and facile integration with other microfabricated devices. Moreover, our device employs concurrent high-sensitivity optical magnetometry to detect the [129]Xe NMR signal. Our results are relevant in light of recent innovations in NMR[14], which include sensitivity enhancement through the use of optically hyperpolarized [129]Xe gas[2] and optical magnetometry[15], and the development of microfluidic NMR methods[16]. Our device combines all of these technologies in a single chip, and may enable the implementation of fully integrated [129]Xe NMR instrumentation in low-cost, portable, and sensitive lab-on-a-chip devices.

## Results

**Microfabricated source of hyperpolarized $^{129}$Xe.** The chip, fabricated from silicon and glass, consists of four chambers 1 mm thick and several millimeters wide connected by micro-channels, as shown in Figure 1a. A fraction of the $^{129}$Xe atoms in the pump chamber (see Figure 1a) becomes polarized through spin exchange with optically pumped $^{87}$Rb atoms at a pumping rate proportional to the $^{87}$Rb polarization, $P_{Rb}$, and the spin-exchange rate, $R_{se} = n_{Rb}\gamma_{se}$, with $n_{Rb}$ being the alkali atomic number density and $\gamma_{se}$ the $^{87}$Rb - $^{129}$Xe spin-exchange rate coefficient[17]. Under gas flow, polarized $^{129}$Xe atoms exit the pump chamber and move through the connecting channel into the probe chamber. Except where stated otherwise, both larger $^{87}$Rb atomic density and stronger laser illumination in the pump chamber ensure that the SEOP rate in the pump chamber is much larger than that in the probe chamber.

**Optical detection of polarized $^{129}$Xe.** Using the ensemble of $^{87}$Rb atoms in the pump and probe chambers as magnetometers[7,18,19] sensitive to magnetic fields along the y axis as defined in Figure 1a, we detected the $^{129}$Xe magnetization, and thus its polarization[20], in each chamber as a function of the experimental conditions. The field $B_{Rb}$ sensed by $^{87}$Rb due to $^{129}$Xe magnetization $M_{Xe}$ is approximately[21]

$$B_{Rb} = \frac{2}{3}\mu_o \kappa_o M_{Xe} \quad (1),$$

where $\mu_o$ is the vacuum permeability, $M_{Xe} = \mu_{Xe} P_{Xe} n_{Xe}$ with $\mu_{Xe}$, $P_{Xe}$, and $n_{Xe}$ being the $^{129}$Xe nuclear magnetic moment, polarization, and atomic density, respectively. Here, $\kappa_o$ represents the enhancement factor due to the Fermi-contact interaction between the valence electron of $^{87}$Rb and the $^{129}$Xe nuclear spin[7,18-21], previously measured to be $\kappa_o \approx$ 500[22]. Measurement of the $^{129}$Xe free-induction decay (FID) in each chamber is made with the magnetometers after rotating the $^{129}$Xe longitudinal polarization along the z-axis into the transverse x-y plane with a short transverse field pulse, as shown in Figure 1c.

Figure 2 demonstrates the efficient transport of polarized $^{129}$Xe from the pump chamber to the probe chamber while the gas is flowing. Figure 2a shows the measured FID in the pump chamber. Figure 2b shows the measured FID in the probe chamber recorded with the pump beam blocked, so that there is no flow of polarized $^{129}$Xe into the probe chamber. Under these conditions we observe a weak probe FID with amplitude of 1 nT, which we attribute to weak SEOP of $^{129}$Xe by the polarized $^{87}$Rb in the probe chamber. Figure 2c shows that the probe FID changes significantly when the pump beam is unblocked; its amplitude and phase follow those of the pump chamber observed in Figure 2a, indicating that the detected $^{129}$Xe was polarized in the pump chamber and transported to the probe chamber by the flow of gas.

**Characterization of $^{129}$Xe polarization.** Next, we characterized the $^{129}$Xe polarization in the pump and probe chambers as a function of total gas flow rate, alkali atomic density, and pumping power. Figure 3a shows the amplitude of the FIDs and the corresponding $^{129}$Xe polarization in the pump and probe chambers, which is estimated using $\kappa_o = 500$[22] in Equation (1), as a function of total gas flow rate. We observe that the polarization in the pump chamber decreases with flow, while in the probe chamber it builds up, peaks at a flow rate at which the influx of polarized $^{129}$Xe is comparable with its spin-relaxation rate, and then follows the pump polarization.

The dependence of the $^{129}$Xe polarization in the pump chamber on the alkali atomic density, without gas flow, is shown in Figure 3b. In our device, relaxation due to $^{129}$Xe collisions with the glass walls is much faster than relaxation from spin-exchange collisions. Thus Equation (2.1) (see Methods) predicts a linear relationship between $^{129}$Xe polarization and the $^{87}$Rb - $^{129}$Xe spin-exchange rate, which is proportional to the $^{87}$Rb atomic density. This relationship, and a constant $^{87}$Rb polarization, can explain the linear behavior of the $^{129}$Xe polarization for low alkali atomic densities. At higher alkali densities the ensemble becomes optically thick and, given a finite amount of pumping power, the alkali polarization decreases. As a result the $^{129}$Xe polarization no longer increases linearly with alkali density. This limitation can be observed in Figure 3c, which shows the $^{129}$Xe polarization in the pump chamber as a function of pumping power for low and high values of the optical depth $OD_o$. At low optical thickness the $^{129}$Xe polarization saturates, whereas for the larger optical depth it remains unsaturated at the highest applied optical powers. These results suggest that larger $^{129}$Xe polarizations can be achieved by increasing both the $^{87}$Rb density and pumping rate.

**Discussion**

The current polarizations achieved in our device are much lower than the near unity polarizations achieved in some large-scale polarizers[10,23]. However, as can be extrapolated from Figures 3b and 3c larger polarizations are in principle achievable in our device. We note that the $^{129}$Xe relaxation observed in our system is 3 to 10 times faster than previously measured in sealed microfabricated cells with 1 mm$^3$ volume[24,25]. In this previous work [129]Xe polarizations on the order of 5% were achieved using similar [87]Rb densities and slightly larger light intensities than those used in our device. These observations suggest room for improvement in future designs and motivate further study of the limitations of SEOP in microfabricated confinement volumes. The combination of higher [87]Rb density, greater optical pumping power, and improved relaxation times may allow for [129]Xe polarizations greater than 5%. The current experiment is designed to produce a large magnetization signal at the expense of polarization fraction through the use of a high [129]Xe partial pressure; the use of lower partial pressures is expected to trade off magnetization signal for increased polarization fraction. An important figure of merit of a polarizer's performance is its spin-transfer efficiency[1]; for our device the spin transfer efficiency is approximately 0.0038, competitive with previous polarizers[10], and only one order of magnitude smaller than the largest possible efficiency. Another figure of merit is the ratio of the polarized atom flow rate to the optical pumping power. We obtain 0.45 µL/(sec-W) for our device, again comparable to this quantity in large-scale polarizers[23].

In this work we employ the [87]Rb atoms as *in situ* detectors[7,18,19] of the [129]Xe magnetization and take advantage of the Fermi-contact interaction[7,18-21] to enhance the detection sensitivity to [129]Xe by a factor of approximately 500. This approach represents an important feature of our device that can be used to detect or monitor the [129]Xe polarization with high sensitivity at low magnetic fields. Alternatively our device can be easily integrated with microfabricated magnetometers[26] placed outside of the pump and probe chambers. Microfabricated magnetometers have demonstrated sensitivities of 5

fT/√Hz over bandwidths of 100 Hz[26], enough to detect the magnetic fields due to the currently produced $^{129}$Xe polarization, and have been used to detect low-field NMR in microfluidics[27]. This approach may be useful for applications that are not compatible with the presence of alkali atoms in the microfluidic channels.

Our device, the first to hyperpolarize $^{129}$Xe atoms in one chamber of a microfabricated device and optically interrogate them in another, will enable several new NMR applications without the need for immobile and expensive laboratory-scale instrumentation. We envision the development of portable high-density microfluidic arrays of xenon biosensors for the analysis of complex mixtures[4,16,28] in confined spaces such as micro-scale chemical reactors, or where access to hyperpolarized $^{129}$Xe produced by large-scale polarizers is impractical, such as in field applications. In such applications, our device would produce hyperpolarized $^{129}$Xe for dissolution into a liquid through use of a microfluidic gas-liquid mixer[29]. The NMR-encoded $^{129}$Xe might be removed from the liquid through a microfluidic analog of a superhydrophobic thin film[30] and then detected *in situ* with our integrated magnetometer, enabling extremely high chemical sensitivity, or detected ex-situ with a physically separate chip-scale atomic magnetometer. Additional experiments suggested for large-scale $^{129}$Xe polarizers[14] might also benefit from implementation on our portable, inexpensive device.

**Methods**

**Experimental setup.** The chip is placed inside a two-layer magnetic shield and attached to a gas manifold (Supplementary S5). Fine control of the gas flow rate through the chip is achieved by a leak valve placed downstream from the chip. Two independently controlled laser beams, tuned to the D1 optical transition of $^{87}$Rb at 794.7 nm, irradiate the pump and probe chambers. A distributed-feedback laser provides the pump laser beam, whereas light produced by a vertical-cavity surface-emitting laser is used as the probe laser beam. For most of the experiments, the optical power of the pump laser beam was 8 mW, while that in the probe was 100 µW. A set of Helmholtz coils is used to provide transverse magnetic fields in the x and y directions and a solenoid is used to provide a longitudinal magnetic field in the z direction. The pump and probe chambers are AC heated using two independent sets of surface mount resistors attached to the cell. The temperature at the input and output chamber is measured using two thermistors that are in close contact with the windows of those chambers. The device operates over a range of temperatures from 120 C to 150 C. The $^{87}$Rb atomic density in each chamber is characterized by measuring the optical absorption of the laser beams as their wavelengths are scanned across the D1 optical line of $^{87}$Rb. From the measured on-resonance optical depth $OD_o$ we calculate the atomic density using $OD_o = \sigma_o n_{Rb} l$ with $l = 1$ mm being the length of the beam path inside the pump chamber and $\sigma_o = 2 r_e c f / \Delta v$ being the on-resonance optical cross section where $r_e = 2.8 \times 10^{-13} cm$ is the classical electron radius, $c$ is the speed of light in vacuum, $f \cong 1/3$ is the oscillator strength of the D1 resonance, and $\Delta v$ =8 GHz is the full-width at half-maximum of the optical line obtained from the fitted optical absorption spectrum.

**Model for the $^{129}$Xe polarization.** The measurements of the $^{129}$Xe polarization in the pump $P_{Xe}^{Pu}$ and probe $P_{Xe}^{Pr}$ chambers can be interpreted in light of a model of the form

$$P_{Xe}^{Pu} = \frac{R_{se}^{Pu}}{R_{se}^{Pu} + R_{wall}^{Pu} + R_{flow}^{Pu}} P_{Rb}^{Pu} \quad (2.1), \quad P_{Xe}^{Pr} = \frac{R_{flow}^{Pr}}{R_{se}^{Pr} + R_{wall}^{Pr} + R_{flow}^{Pr}} P_{Xe}^{Pu} \quad (2.2),$$

where the superscripts Pu and Pr indicate the pump and probe chambers, respectively, $R_{wall}$ is the $^{129}$Xe spin-destruction rate due to collisions with the chamber walls and $R_{flow}$ is the inverse of the $^{129}$Xe transit time through the chamber. These equations are useful for understanding some characteristics of our device and guide future designs. We note though, that they neglect diffusion, which limits the estimates of Equation (2.2) at low flow rates. For example, Figure 3a demonstrates that, when no gas is flowing, the polarization in the probe chamber is larger when the pump laser is on than when it is off. We have performed experiments indicating that this is due to the transport of polarized $^{129}$Xe through diffusion (Supplementary S1).

**Atomic magnetometer response and FID estimates**. The optically pumped $^{87}$Rb magnetometers utilized here are very similar to those discussed in references 18 and 19. The magnetometers are implemented by introducing a radio-frequency field with 7 kHz modulation frequency and 1μT amplitude along the y axis. The signals of the magnetometers are extracted by lock-in detection of the transmitted laser intensity at the modulation frequency. The measured noise floors for the pump and probe magnetometers were 5 $pT/\sqrt{Hz}$ and 16 $pT/\sqrt{Hz}$, respectively, with a bandwidth of 40 Hz that is limited by the response of the lock-in amplifiers. This corresponds to a single-shot detection limit of a polarized ensemble of ~5×10$^{11}$ $^{129}$Xe atoms in each chamber. On resonance, the magnetometers respond linearly to magnetic fields along the y axis that are smaller than the resonance line-width ($\Delta v_{HWHM} \approx 1\ \mu T$ at zero-light levels dominated by collisions with Xe atoms, and twice or larger with optical pumping). For each change in the device parameters, the linear response of the magnetometers was obtained by extracting the on-resonance slope of the magnetic response acquired by scanning the magnetic field along the y axis. We then used this slope to calibrate the amplitude of the FIDs in magnetic field units. We fit the FID signal in the pump chamber to a single exponentially decaying sinusoid, from which we extracted the amplitude, decay time constant, and Larmor precession frequency of $^{129}$Xe. We obtained estimates for the FID amplitude in the probe chamber by measuring the amplitude of the first peak and valley in the probe FID signal. Because transverse polarization is transferred from pump to probe chambers as the FID signals are being taken, nonexponential decay of the probe FID can occur, preventing the simple analysis of the data with a single time constant.

**Acknowledgments**

The authors thank S. Schima for help with the fabrication of the device, E. Pratt and K. Stupic for useful comments on the manuscript, A. Pines for helpful discussions, and L. Jiménez for help in the preparation of Figure 1. This work is a contribution of the National Institute of Standards and Technology (NIST), an agency of the U.S. government, and is not subject to copyright. Research was supported by the U.S. Department of Energy, Office of Basic Energy Sciences, Division of Materials Science and Engineering under contract No. DE-AC02-05CH11231 (DJK, SJS, HLR, and VSB). R.J.M. was supported in part by the Roberto Rocca Education Program.

**Author contributions**

All authors contributed extensively to the work presented in this paper.

**Additional information**

Supplementary information is available in the online version of the paper.

Correspondence should be addressed to V.S.B. and J.K.

**Competing financial interests**

The authors declare no competing financial interests.

**Figure 1. The microfluidic chip $^{129}$Xe polarizer.** (a) A gas mixture containing 400 Torr N$_2$ and 200 Torr Xe in natural isotopic abundance (26.4% $^{129}$Xe content) flows from a bulk gas manifold into the inlet chamber, through the pump and probe chambers, and out of the outlet chamber. The chip is loaded with 2 mg of $^{87}$Rb metal (Supplementary S5). Unpolarized $^{129}$Xe atoms entering the pump chamber become polarized through spin exchange with optically pumped $^{87}$Rb. The $^{129}$Xe then moves downstream, passes through a microchannel into the probe chamber, and eventually exits the device through the output chamber. Optical characterization of the $^{129}$Xe polarization in the pump and probe chambers is carried out using the ensemble of $^{87}$Rb atoms in each chamber as *in situ* magnetometers. (b) The silicon chip footprint is 3 cm × 1 cm, with a thickness of 1 mm. The dimensions of the pump and probe chambers are 5 mm × 5 mm × 1 mm and 3 mm × 3mm × 1 mm, respectively, whereas the channel connecting the pump and probe chambers is 1 mm x 0.3 mm x 0.3 mm. Two tall, narrow grooves are etched from the middle of the chip to provide thermal isolation between the two sides of the device. (c) Pumping and probing sequence for $^{129}$Xe. Pumping is carried out continually in the pump chamber in the presence of a longitudinal field of $B_z = 0.8$ µT. Every 10-20

seconds, a transverse field of magnitude 5.3 µT is switched on for 4 ms to tip the $^{129}$Xe atoms onto the x-y plane and initiate the $^{129}$Xe precession about the longitudinal axis.

**Figure 2. Transport of polarized $^{129}$Xe.** (a) Free-induction decay (FID) of the transverse $^{129}$Xe polarization in the pump chamber at a total gas flow rate of 3 µL/s. (b) FID of polarized $^{129}$Xe in the probe chamber with the pump light turned off. (c) FID of polarized $^{129}$Xe in the probe chamber with the pump light turned on. Note that for this particular experiment the FID obtained in (b) is 180 degrees out of phase with that obtained in the pump chamber (a), due to the use of opposite-handed polarizing photons in the pump and probe chambers.

**Figure 3. Device characterization.** (a) $^{129}$Xe free-induction decay (FID) amplitude as a function of total gas flow rate in the pump chamber (blue circles), in the probe chamber with the pump light on (red squares), and with the pump light off (black squares). The solid line corresponds to a fit of the data using Equation (2.1) with $R_{se}^{Pu}$ as a free parameter and with $P_{Rb}^{Pu} = 0.36$ and $R_{wall}^{Pu} = (1.6\ s)^{-1}$ (for details see Supplementary S3). From the obtained value for $R_{se}^{Pu} = 0.012\ s^{-1}$ and for a $^{129}$Rb atomic density of $6 \times 10^{13}$ atoms-cm$^{-3}$ in the pump chamber, which is estimated from the measured on-resonance optical depth (Methods), we estimate $\gamma_{se} = 1.7 \times 10^{-16}$ cm$^3$/s, in agreement with reported values for the spin-exchange rate coefficient due to $^{87}$Rb-$^{129}$Xe binary-collisions[17]. The dashed line corresponds to the evaluation of Equation (2.2) using the fitted value for $P_{Xe}^{Pu}$ and $R_{wall}^{Pr} = (1.6\ s)^{-1}$; this expression ignores $^{87}$Rb - $^{129}$Xe spin exchange and neglects the effects of diffusion, which can explain the discrepancy

between the estimated line and the data at low flow rates. (b) Pump FID amplitude with no gas flow as a function of $^{87}$Rb atomic density and (c) as a function of optical pumping power for $n_{Rb}= 6 \times 10^{13}$ atoms-cm$^{-3}$ (blue circles) and $n_{Rb} = 1 \times 10^{13}$ atoms-cm$^{-3}$ (black triangles). In (c) the solid lines correspond to fits to the data using Equation (1) and Equation (2.1) with $R_{se}^{Pu}$ as a fitting parameter and using $R_{wall}^{Pu} = (1.6 \text{ s})^{-1}$. The dependence of the Rb polarization was extracted from a fit to the measured shift in the $^{129}$Xe Larmor frequency due to the $^{87}$Rb magnetization as a function of pumping power (Supplementary S3).

**Figure 1.**

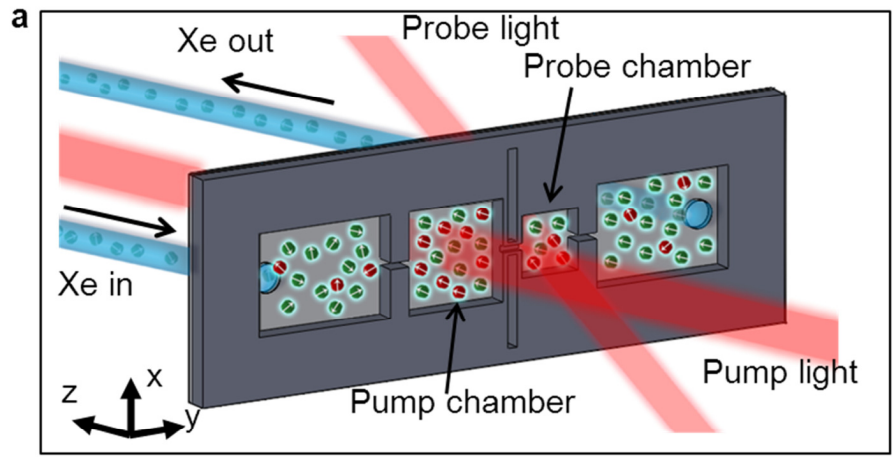

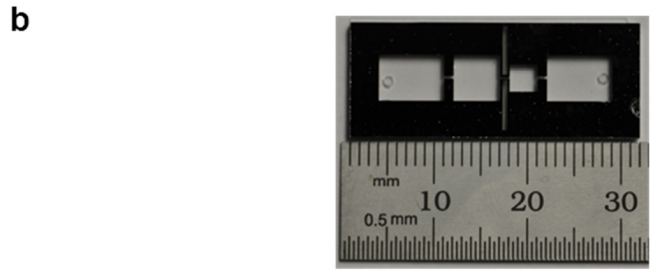

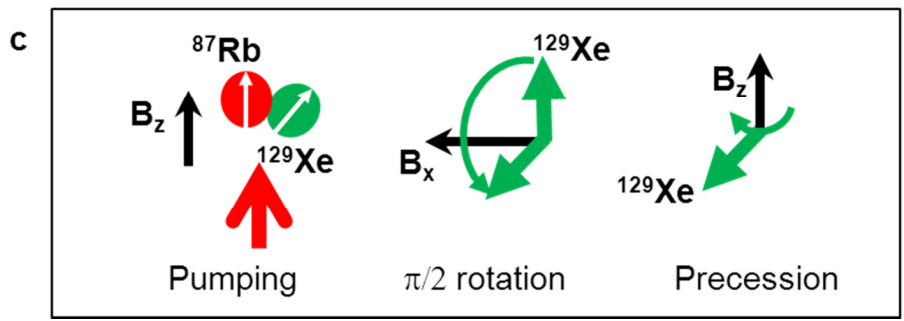

**Figure 2.**

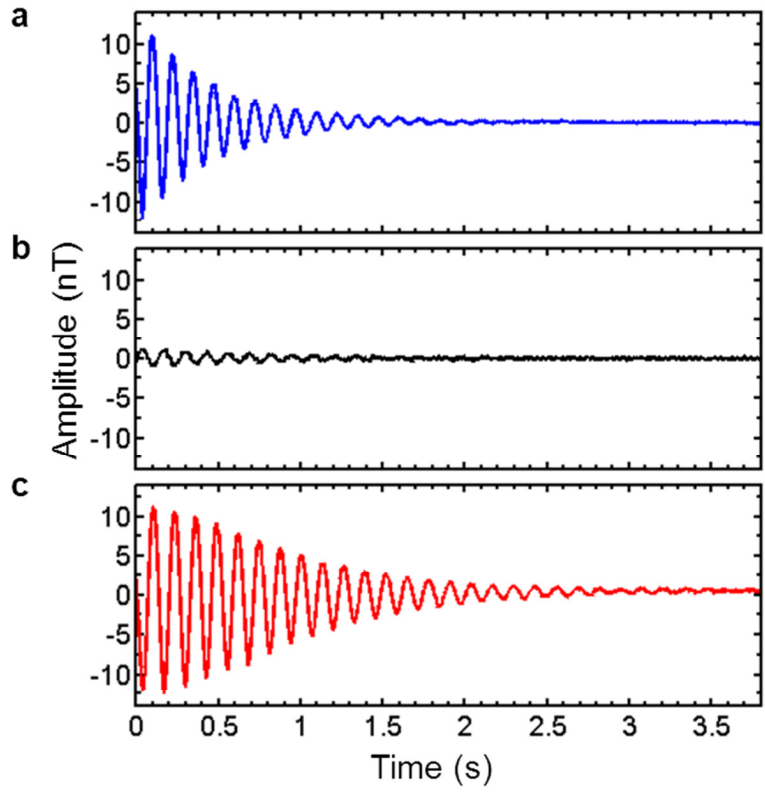

**Figure 3.**

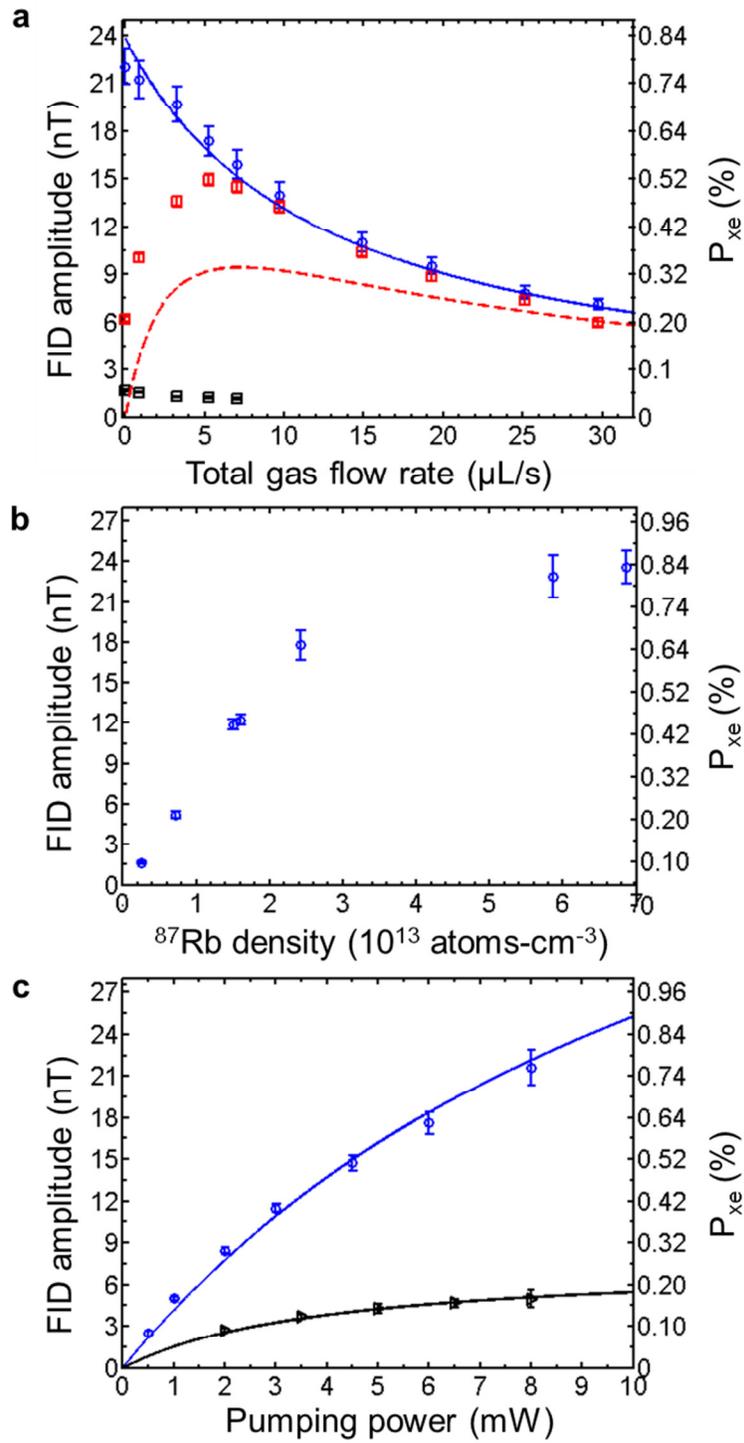